\documentclass[aps,prl,twocolumn,floatfix,superscriptaddress]{revtex4}
\usepackage{amsmath}
\usepackage{amssymb}
\usepackage{graphicx}
\usepackage{dcolumn}
\usepackage{bm}

\begin{document}

\title{Optimal Control of Quantum Rings by Terahertz Laser Pulses}

\author{E. R{\"a}s{\"a}nen}
\email[Electronic address:\;]{esa@physik.fu-berlin.de}
\affiliation{European Theoretical Spectroscopy Facility (ETSF),
Institut f{\"u}r Theoretische Physik, Freie Universit{\"a}t Berlin, 
Arnimallee 14, D-14195 Berlin, Germany.}
\author{A. Castro}
\affiliation{European Theoretical Spectroscopy Facility (ETSF),
Institut f{\"u}r Theoretische Physik,  
Freie Universit{\"a}t Berlin, Arnimallee 14, D-14195 Berlin, Germany.}
\author{J. Werschnik}
\affiliation{European Theoretical Spectroscopy Facility (ETSF), 
Institut f{\"u}r Theoretische Physik,  
Freie Universit{\"a}t Berlin, Arnimallee 14, D-14195 Berlin, Germany.}
\author{A. Rubio}
\affiliation{Departamento de F{\'i}sica de Materiales, Facultad de Qu{\'i}micas
Universidad del Pa{\'i}s Vasco, Centro Mixto CSIC-UPV, Donostia International
Physics Center (DIPC), E-20018 Donostia-San Sebasti{\'a}n, Spain}
\affiliation{European Theoretical Spectroscopy Facility (ETSF),
Institut f{\"u}r Theoretische Physik,  
Freie Universit{\"a}t Berlin, Arnimallee 14, D-14195 Berlin, Germany.}
\author{E. K. U. Gross}
\affiliation{European Theoretical Spectroscopy Facility (ETSF),
Institut f{\"u}r Theoretische Physik,  
Freie Universit{\"a}t Berlin, Arnimallee 14, D-14195 Berlin, Germany.}


\begin{abstract}
Complete control of single-electron states in a 
two-dimensional semiconductor quantum-ring model is established,
opening a path into coherent laser-driven single-gate qubits. 
The control scheme is developed in the framework of optimal 
control theory for laser pulses of two-component polarization. 
In terms of pulse lengths and target-state occupations, 
the scheme is shown to be superior to conventional 
control methods that exploit Rabi oscillations
generated by uniform circularly polarized pulses.
Current-carrying states in a quantum ring can be used to
manipulate a two-level subsystem at the ring center. 
Combining our results, we propose a realistic approach to
construct a laser-driven single-gate qubit that has
switching times in the terahertz regime.
\end{abstract}

\maketitle

In recent years there has been wide interest in
quantum control of nanoscale systems.
One of the main motivations behind these studies 
arises from the possibilities 
of using tailored laser-pulse sequences for 
logic operations~\cite{computer}. 
Semiconductor quantum dots and quantum rings (QRs)~\cite{ringreview} 
are likely to play an important role in these far-reaching 
developments. Their atomlike properties together
with a high flexibility in size and shape 
construct an ideal playground for quantum control. 

A fundamental question in laser control of a general $N$-level 
quantum system is {\em controllability}, i.e., if the
control target such as a certain eigenstate can be 
reached even in principle. For example, 
full population transfer in a {\em parabolic} quantum dot 
is not possible (in dipole approximation)~\cite{qdcontrol}. 
The first steps towards laser control of QRs were recently taken by 
Pershin and Piermarocchi~\cite{pershin}, who theoretically
analyzed the formation of currents through population transfer 
in narrow QRs subjected to a circularly polarized laser pulses. 
Similar approach has been used also to generate ring currents in 
circular biomolecules such as Mg--porphyrin~\cite{manz}.
Quantum optimal control theory (OCT)~\cite{kosloffandpeirce}
is a powerful tool to find optimal laser pulses for controlling a quantum 
system. The iterative scheme developed within 
OCT~\cite{zhu} converges monotonically to an optimal laser pulse 
for reaching the
prescribed target, 
such as a desired final quantum state, at the end of the 
pulse. 

In this Letter we apply OCT to semiconductor
QRs of finite ring width. We
construct {\em optimal} two-component 
laser pulses that drive the QR from a given initial state
to any predefined target state.
These terahertz pulses generate the desired transitions 
in significantly shorter times and higher accuracies than previously
used circularly polarized continuous waves (CWs) of finite lengths. 
Finally we sketch how the full control of current-carrying states
in QRs enables the construction of a coherent laser-driven single-gate qubit.

\begin{figure}
\includegraphics[width=0.85\columnwidth]{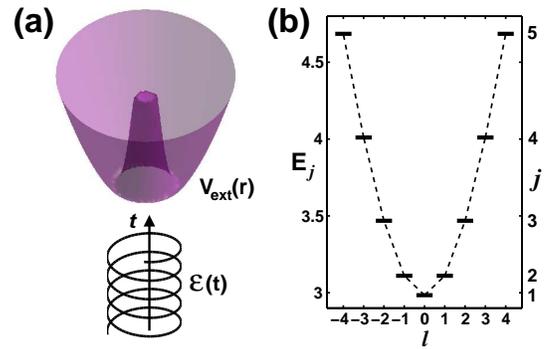}
\caption{(a) Shape of the external confining potential 
for a quantum ring and an example of a circularly polarized
laser field. (b) Energy-level spectrum of a quantum ring.
The transitions are allowed along the dashed line so that 
$\Delta l=\pm 1$.}
\label{vext}
\end{figure}

The time-evolution of our system is described by 
the time-dependent Schr{\"o}dinger equation
\begin{equation}
i\hbar\frac{\partial}{\partial t}\Psi({\mathbf r},t)=
\left[{\hat H}_0-{\hat {\mathbf \mu}}{\boldsymbol \epsilon}(t)\right]
\Psi({\mathbf r},t),
\end{equation}
where  
${\boldsymbol \epsilon}(t)=(\epsilon_x(t),\epsilon_y(t))$
is the two-component laser field propagating
in {\em z} direction. 
The interaction between the field and the electron 
is modeled in the dipole approximation
(length gauge) with the dipole operator 
${\hat {\mathbf \mu}}=-e\mathbf{r}$. 
The static effective Hamiltonian 
for the semiconductor QR located on the {\em xy} plane 
is written as 
\begin{equation}
{\hat H}_0=-\frac{\hbar^2}{2m^*}\nabla^2+
\frac{1}{2}m^*\omega_0^2r^2 + V_0 e^{-r^2/d^2},
\label{static}
\end{equation}
where the potential part with $r^2=x^2+y^2$ consists of a parabolic 
confinement and a Gaussian peak located at the center.
Together they determine the shape of a two-dimensional 
QR which is visualized in Fig.~\ref{vext}.
We choose the parameter values $\hbar\omega_0=10$ meV, 
$V_0=200$ meV, and $d=10$ nm in order to follow the
energy and length scales of QR experiments~\cite{ringreview}. 
The parameters
yield a ring radius of $r_0\approx 22$ nm.
The effective mass of the electron in GaAs semiconductor
medium is $m^*=0.067m_e$. Using the dielectric constant
$\kappa=12.7\epsilon_0$ for GaAs, the effective atomic
units [marked as (a.u.) in the results below] scale as
${\rm Ha}^*=(m^*/\kappa^2){\rm Ha}\approx 11.30$ meV, 
$a_B^*=(m^*/\kappa)a_0\approx 10.03$ nm, and 
$u^*_t=\hbar/{\rm Ha}^*\approx 58.23$ fs.

Figure~\ref{vext}(b) shows the energy-level spectrum
for nine lowest states $\psi^l_j$, where $l$ is the
angular momentum and index $j=|l|+1$ enumerates the
levels in energy. Due to the finite width of the ring,
there are other radial bands which, however, appear
at higher energies, e.g., the second band appears
at $E=5.03$. The selection rules allow
transitions only between states with consecutive 
angular momenta $l=\ldots,-2,-1,0,1,2,\ldots$, i.e.,
along the dashed line marked in the figure. 
The transition probabilities between the levels
can be calculated as 
$P_{jk}=|\left<\psi_k|\hat{\mu}|\psi_j\right>|^2$.
For the first levels shown in Fig.~\ref{vext}(b)
we get relative transition probabilities 
$P_{12}=1$, $P_{23}=1.04$,  $P_{34}=1.12$, and 
$P_{45}=1.24$. Hence, the two-dimensional nature
of the ring is reflected also in the varying $P_{jk}$:
in one-dimensional rings, the dipole matrix elements 
are constant along the transition lines~\cite{pershin}. 

We apply OCT to find optimal laser pulses
for population transfer from the initial 
state $\Phi_i=\Psi(t=0)$ to the target state $\Phi_f$, so 
that both of these states are preselected from 
the QR eigenstates $\psi^l_j$. In OCT 
formalism we thus use the projection operator 
$|\Phi_f\big>\big<\Phi_f|$ as the target operator, whose
expectation value is maximized at the end of the pulse ($t=T$). 
This corresponds to maximizing the overlap
\begin{equation}
J_1[\Psi]
=\left<\Psi(T)|\Phi_f\right>
\left<\Phi_f|\Psi(T)\right>
=|\left<\Psi(T)|\Phi_f\right>|^2.
\end{equation}
We also require that the fluence (time-integrated intensity)
of the laser pulse is as small as possible
by minimizing the functional
\begin{equation}
J_2[\epsilon]=-\int_0^T dt\,\alpha(t)\left[\epsilon_x^2(t)+\epsilon_y^2(t)\right], 
\label{field}
\end{equation}
where the predefined function $\alpha(t)$ acts as a penalty
factor, which allows us to impose constraints on the envelope
function of the laser pulse~\cite{sundermann}. We can apply, 
e.g., sinusoidal envelope functions as demonstrated in 
Ref.~\onlinecite{janthesis}. However, in the examples below 
we have set $\alpha=1$ (rectangular pulses) in order
to simplify the comparison between OCT and CW approaches.
As the second constraint, the wave function and its complex 
conjugate need to satisfy
the time-dependent Schr\"odinger equation. This condition is 
expressed in a functional form (in a.u.) as
\begin{equation}
J_3[\epsilon,\Psi,\chi]=-2\,{\rm Im}\,\int_0^T dt\,
\left<\chi(t)\left|\left(i\partial_t-\hat{H}(t)\right)\right|\Psi(t)\right>, 
\end{equation}
where $\chi(t)$ is the Lagrange multiplier.
Now, the minimum of the total Lagrange functional 
$J=J_1+J_2+J_3$
can be determined by setting the variations with respect to 
$\Psi$, $\epsilon$, and $\chi$ independently to zero. This yields a
set of control equations 
\begin{eqnarray}
i\partial_t\Psi(t) & = & \hat{H}\Psi(t), \,\,\Psi(0)=\Phi_i, \\
i\partial_t\chi(t) & = & \hat{H}\chi(t), \,\,\chi(T)=\Phi_f\left<\Phi_f|\Psi(T)\right>,\\
\epsilon(t) & = & -\frac{1}{\alpha}{\rm Im}\,\left<\chi(t)|\hat{\mu}|\Psi(t)\right>,
\end{eqnarray}
which can be solved iteratively in order to find
the optimized field $\epsilon(t)$ at the end of the procedure~\cite{zhu}.
For the forward and
backward propagation of $\Psi(t)$ and $\chi(t)$ we apply a rapidly
converging scheme discussed in Refs.~\onlinecite{zhu,janthesis}.
This computational scheme has been recently included in the {\tt octopus}
code~\cite{octopus}.

\begin{figure}
\includegraphics[width=0.9\columnwidth]{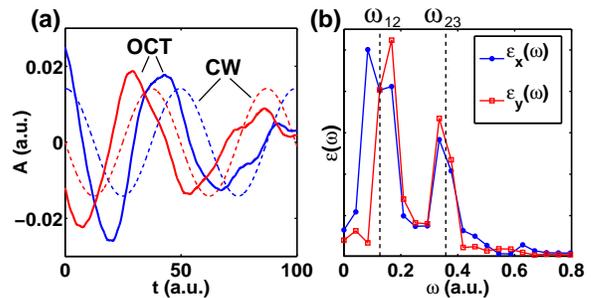}
\caption{(Color online) (a) Components $\epsilon_x(t)$ (blue) and 
$\epsilon_y(t)$ (red) of the optimized (solid lines) and 
uniform laser pulses (dashed lines) for 
$|1\big>\rightarrow|2\big>$ transition. (b) Spectrum of the
optimized field. The dashed lines correspond to eigenfrequencies
$\omega_{12}$ and $\omega_{23}$.}
\label{laser}
\end{figure}

The advantage of OCT becomes obvious already in the 
simplest transition from $\psi_1^{0}$ to $\psi_2^{1}$ (denoted
below as $|1\big>\rightarrow|2\big>$).
Figure~\ref{laser}(a)
shows the {\it x} and {\it y} components of the optimized pulse
when the pulse length is fixed to $T=100$ ($\sim 5.8$ ps) and
the penalty factor is $\alpha=1$. In comparison, the dashed lines 
show the CW:
$\epsilon_{\rm CW}(t)=A\left[\cos(\omega t)
{\hat x} \pm \sin(\omega t){\hat y} \right]$,
where ${\pm}$ denotes $\sigma_{\pm}$ polarization, and
$A$ and $\omega$ are the amplitude and frequency of the field, 
respectively. Here we choose $\omega=\omega_{12}$ and
$A=\Omega_R/\mu_{12}=\pi/\mu_{12}T$, where $\Omega_R$ is the
Rabi frequency. These parameters correspond to the
$\pi$-pulse condition.
The CW leads to an occupation of $0.99$ 
in the second state which, as demonstrated below, is 
insufficient in terms of the controllability of the QR. 
In contrast, the optimized pulse produces a value $0.9998$ 
for the same pulse length. The underlying reason for
the difference becomes evident from the spectrum of
the optimized pulse shown in Fig.~\ref{laser}(b).
Namely, the pulse contains a considerable fraction 
of the next eigenfrequency $\omega_{23}$, 
preventing the population flow into the third state.
If the system is subjected to a CW instead,
the occupation loss is inevitable, and becomes 
pronounced at shorter pulse lengths when the amplitudes 
are correspondingly higher. 

\begin{figure}
\includegraphics[width=0.85\columnwidth]{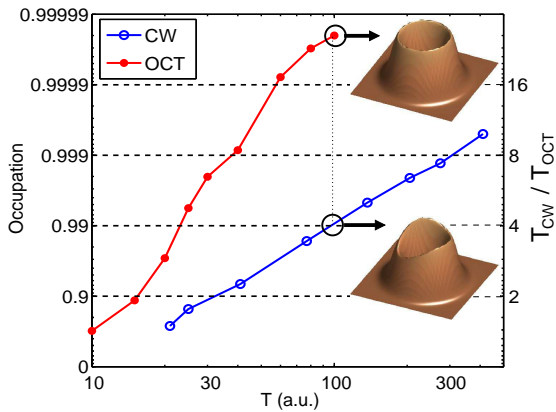}
\caption{(Color online) Maximum occupation of the target state in transition 
$|1\big>\rightarrow|2\big>$ as a function of the pulse length.
The blue open circles correspond to continuous waves and red filled circles
to the optimal-control result. The insets show the
densities $|\Psi(T=100)|^2$ when the corresponding achieved occupations 
are 0.99 and 0.9998 for these pulse types, respectively.}
\label{speed}
\end{figure}

In Fig.~\ref{speed}
the yields of OCT and CW approaches 
for the $|1\big>\rightarrow |2\big>$ transition are compared 
for different pulse lengths in a logarithmic scale. 
Considering the
required pulse lengths for desired occupation 
accuracies, OCT is clearly superior to the CW approach.
The relation between the pulse lengths of
these methods $T_{\rm CW}/T_{\rm OCT}$ for a certain 
occupation always doubles when the accuracy
is increased by an order of magnitude. 
For example, the occupation of $0.99$ 
requires $T_{\rm CW}\sim 100$ or $T_{\rm OCT}\sim 25$,
whereas for an accuracy of $0.999$ the CW needs to be increased up to
$T_{\rm CW}\sim 300$ and the optimized pulse only to
$T_{\rm OCT}\sim 35$.
The main benefit in very high occupations
achieved by short optimized pulses is the ability to 
perform a large number of successive operations 
(see below). In {\em single} transitions we can qualitatively 
assess the required occupation accuracies by checking the 
final densities of the propagated wave function. 
The insets of Fig.~\ref{speed} reveal that occupations 
close to $0.999$ are required for the target states in
order to preserve a reasonable circular symmetry of 
the wave function. 

The benefits of OCT become even more pronounced
in multilevel transitions. As shown above,
there is only a single excitation path in the lowest 
radial band of our QR. The path is subjected to strict
selection rules that allow transitions only
between consecutive levels with $\Delta l=\pm 1$.
Hence, in the CW scheme the intermediate states require 
full population before occupation of the final state. 
Eventually this leads to very slow transition 
processes sensitive to external perturbations of
the real device. In OCT approach instead,
full population of the intermediate states is not needed,
since the components in the optimized pulse drive
the occupation continuously toward the target state,
still obeying the selection rules. 
For example, in $|1\big>\rightarrow |4\big>$ transition
the maximum occupation required for the intermediate 
states $|2\big>$ and $|3\big>$ is $\sim 60\ldots 80 \%$,
depending on the pulse length (here $T=50\ldots 200$).

\begin{figure}
\includegraphics[width=0.9\columnwidth]{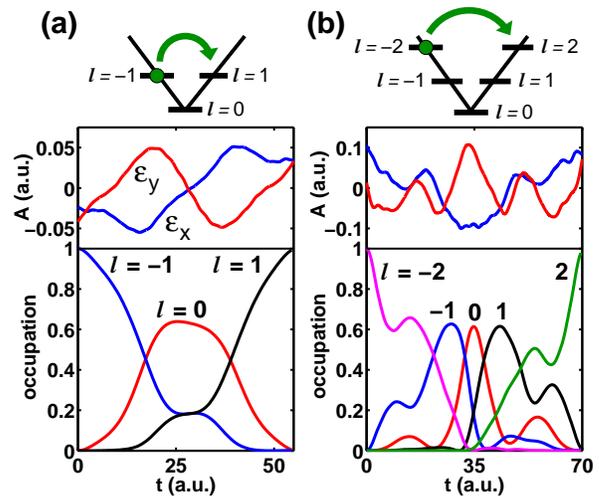}
\caption{(Color online) Schematic picture of transitions
from $l=-1$ to $l=1$ (a) and from $l=-2$ to $l=2$ (b) (upper panel),
optimized fields for these transitions (middle panel), and the
occupations of the states (lower panel).}
\label{finalqubit}
\end{figure}

The chirality of a QR state can be changed in 
multilevel transitions from
$l$ to $-l$ (or vice versa). Figure~\ref{finalqubit}
shows schematic pictures (upper panel) of transitions 
from $\psi_2^{-1}$ to $\psi_2^{1}$ (a) and from
$\psi_3^{-2}$ to $\psi_3^{2}$ (b). The optimized pulses
given in the middle panel of Fig.~\ref{finalqubit} are 
approximately symmetric around the midpoints at $t=T/2$, and
at this point the polarization direction of the field changes 
from $\sigma_{+}$ to $\sigma_{-}$. Hence, a backward process
can be obtained by applying a pulse where the {\it x} and {\it y} 
components are exchanged. 

The intermediate states, which in this case contain the ground 
state with $l=0$, require only partial occupation in the transition 
process. The intermediate occupations shown in the lower panel of 
Fig.~\ref{finalqubit} reach a maximum of $\sim 60 \%$ 
in both $l=-1\rightarrow 1$ and $l=-2\rightarrow 2$ transitions.
Similar to the pulse shapes, the occupation curves are approximately 
symmetric around $t=T/2$. The final occupations
of the target states are $0.998$ and $0.996$, respectively. 

\begin{figure}
\includegraphics[width=0.75\columnwidth]{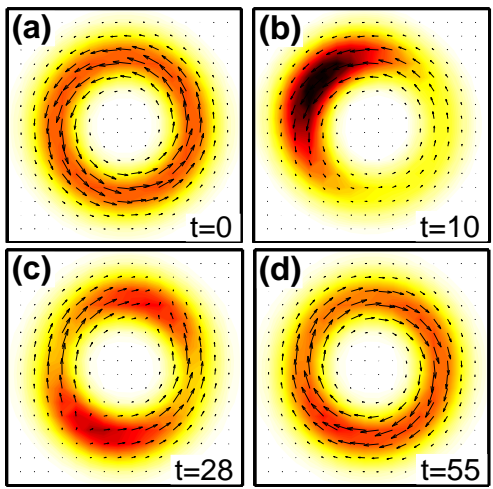}
\caption{(Color online) Time-dependent electron density 
$|\Psi(\mathbf{r},t)|^2$ [light shade (yellow), low; black, high]
and the current $\mathbf{j}(\mathbf{r},t)$ (arrows) 
at different times in the transition
from $l=-1$ to $l=1$ driven by the optimized
pulse [see Fig.~\ref{finalqubit}(a)].}
\label{qubit}
\end{figure}

In Fig.~\ref{qubit} 
we plot the time-dependent electron densities $|\Psi(\mathbf{r},t)|^2$ (colorscale)
and the current densities
$\mathbf{j}(\mathbf{r},t)=i\hbar(\Psi\nabla\Psi^*-\Psi^*\nabla\Psi)/2m^*$ (arrows)
at different times in the transition from $l=-1$ to $l=1$,
corresponding to Fig.~\ref{finalqubit}(a).
The circularly symmetric initial state in (a) carries a 
counterclockwise current of $I\approx 0.11\,\mu$A. During the propagation,
the density and current localize first to a single maximum (b). 
At the midpoint (c) there are two density maxima along the ring and
opposite currents between them along the two paths, arising from
the partial occupations of $l=\pm 1$. The net
current is zero at this point. As expected, the final state 
reached at $T=55$ (d) has a symmetric clockwise 
current. 

The circular currents in QRs induce magnetic fields through 
the ring in accordance with the Biot-Savart law. 
The current-flipping processes correspond 
to changes in magnetic fields as $B\hat{z}\leftrightharpoons
-B\hat{z}$ at the center of the ring, where $B=\mu_0 I/2r_0 \approx 3\,\mu$T
for a single electron occupying the state $l=\pm 1$. 
If needed, higher fields can be obtained by occupying higher
states, by decreasing the ring size, or by changing the
semiconductor host material. 

Combining the results presented above, we suggest the following
single-qubit gate construction.
First, the accurate experimental
shape of the QR and thus the static Hamiltonian (\ref{static}) have to be
determined in order to allow for a full control. This can be done, e.g.,
by tuning the potential parameters to reproduce the single-electron 
spectrum obtained in a transport measurement~\cite{jens}. 
Secondly, the QR is controlled 
by a sequential {\em master pulse} achieving first
the initial excitation from $l=0$ to $l=l_m$ (qubit initialization) 
and then performing $l_m\leftrightharpoons -l_m$ operations at desired times so
that the opposite operation is obtained by exchanging
$\epsilon_x$ and $\epsilon_y$ components of the current-flipping pulse. 
The magnetic field induced 
at the ring center changes the spin state (up/down) of
an applicable subsystem, e.g., a magnetic particle~\cite{pershin}
or an attached quantum dot. 
This is feasible within our picosecond-scale gate operation times,
since the characteristic relaxation times are of the order of 
nanoseconds~\cite{pershin}, and the decoherence times 
in quantum dots (of similar physical parameters) have been measured to
be on the millisecond scale~\cite{engel}. 
Actual quantum computation would
still require a two-qubit gate and a stable read-out scheme, but
these developments, as well as detailed experimental considerations 
are beyond the scope of this study.
However, we point out that terahertz frequency
regime is routinely reached by, e.g., quantum cascade lasers, and the
technology of pulse refinement is under rapid development~\cite{tonouchi}.

In conclusion, we have shown that a complete control of
realistic single-electron quantum rings can be obtained using
(quantum) optimal control theory. The optimized pulses perform the 
desired operations at significantly shorter times and 
better accuracies than the continuous waves, and in multilevel
transitions they are not restricted by the need to fully populate 
the intermediate states. This enables a simple construction of a pulse that
coherently flips the spin of a subsystem at the center 
of the ring at preselected times.

\begin{acknowledgments}
This work was supported by the EU's Sixth Framework 
Programme through the Nanoquanta Network of Excellence 
(NMP4-CT-2004-500198), SANES project (NMP4-CT-2006-017310),
DNA-NANODEVICES (IST-2006-029192), BSC (Barcelona Mare 
Nostrum Center), the 2005 Bessel research award of the 
Humboldt Foundation, the Academy of Finland, the Finnish 
Academy of Science and Letters through the Viljo, 
Yrj{\"o} and Kalle V{\"a}is{\"a}l{\"a} Foundation,
and by the Deutsche Forschungsgemeinschaft.
\end{acknowledgments}

\end{document}